\PassOptionsToPackage{unicode}{hyperref}
\PassOptionsToPackage{hyphens}{url}
\documentclass[
  11pt,
  a4paper,
]{article}
\usepackage{amsmath,amssymb}
\usepackage{iftex}
\ifPDFTeX
  \usepackage[utf8]{inputenc}
  \usepackage{textcomp} 
\else 
  \usepackage{unicode-math} 
  \defaultfontfeatures{Scale=MatchLowercase}
  \defaultfontfeatures[\rmfamily]{Ligatures=TeX,Scale=1}
\fi
\ifPDFTeX\else
\fi
\IfFileExists{upquote.sty}{\usepackage{upquote}}{}

\makeatletter
\@ifundefined{KOMAClassName}{
  \IfFileExists{parskip.sty}{%
    \usepackage{parskip}
  }{
    \setlength{\parindent}{0pt}
    \setlength{\parskip}{6pt plus 2pt minus 1pt}}
}{
  \KOMAoptions{parskip=half}}
\makeatother
\usepackage{xcolor}
\usepackage[margin=0.87in]{geometry}
\usepackage{setspace}
\usepackage{longtable,booktabs,array}
\usepackage{calc} 
\usepackage{etoolbox}
\makeatletter
\patchcmd\longtable{\par}{\if@noskipsec\mbox{}\fi\par}{}{}
\makeatother
\IfFileExists{footnotehyper.sty}{\usepackage{footnotehyper}}{\usepackage{footnote}}
\makesavenoteenv{longtable}
\usepackage{graphicx}
\makeatletter
\def\maxwidth{\ifdim\Gin@nat@width>\linewidth\linewidth\else\Gin@nat@width\fi}
\def\maxheight{\ifdim\Gin@nat@height>\textheight\textheight\else\Gin@nat@height\fi}
\makeatother
\setkeys{Gin}{width=\maxwidth,height=\maxheight,keepaspectratio}
\makeatletter
\def\fps@figure{htbp}
\makeatother
\ifLuaTeX
\usepackage[bidi=basic]{babel}
\else
\usepackage[bidi=default]{babel}
\fi
\babelprovide[main,import]{american}

\def\languageshorthands#1{}
\ifLuaTeX
  \usepackage{selnolig}  
\fi
\usepackage[round,authoryear]{natbib}
\setcitestyle{aysep={,},citesep={;}}
\usepackage{bookmark}
\IfFileExists{xurl.sty}{\usepackage{xurl}}{} 
\hypersetup{
  pdftitle={Statistical Inference with LLM-Generated Synthetic Survey Data: A Review of Methods, Challenges, and Future Directions},
  pdfauthor={Yilong Hou and Xiaoning Wang},
  pdflang={en-US},
  hidelinks,
  pdfcreator={LaTeX via pandoc}}

\title{Statistical Inference with LLM-Generated Synthetic Survey Data: A Review of Methods, Challenges, and Future Directions}
\author{
  \begin{minipage}{0.95\textwidth}
  \centering\setstretch{1.15}
  Yilong Hou and Xiaoning Wang\textsuperscript{*}\\[5pt]
  \normalsize School of Data Science and Media Intelligence, Communication University of China\\
  No. 1 Dingfuzhuang East Street, Chaoyang District, Beijing 100024, China\\[5pt]
  \textsuperscript{*}Xiaoning Wang (corresponding author), \href{mailto:sdwangxiaoning@foxmail.com}{sdwangxiaoning@foxmail.com}
  \end{minipage}
}
\date{}

\begin{document}
\maketitle

\begin{abstract}

Large language models can generate answers to survey questions at scale. We review the conditions under which these answers can inform inference about human populations. We compare direct use of generated responses, calibration to population profiles, and estimation that corrects model predictions using observed human outcomes. The review shows how errors in profile composition and conditional responses can offset each other in a population mean while distorting subgroup differences. Repeated generation reduces Monte Carlo error. Confidence intervals for human-population parameters also require a basis for estimating errors in the generated responses. In a paired reanalysis of two public benchmarks, the frequency with which GPT-4 effect estimates exceed human estimates in absolute value is 68.4\% in scenario replications and 52.9\% in text-based survey experiments. These results illustrate how the evidence needed to use synthetic data changes with the estimand. We propose validation designs for subgroup contrasts, external transfer, and comparisons of model-assisted estimators under equal survey budgets.

\end{abstract}


\section{Introduction}\label{introduction}

A large social survey defines a target population, contacts respondents,
measures outcomes, and estimates population parameters from those
measurements. Large language models (LLMs) change how survey-like data
can be produced: they answer items for specified respondent profiles or
report a distribution over options directly. Researchers can elicit many
responses to the same questionnaire, vary profiles, wording, or
experimental materials, and compare the outputs of several models.
Grossmann et al. \citeyearpar{grossmann} and Bail \citeyearpar{bail}
describe how generative AI can enter social science workflows, surveys,
and experiments. Dillion et al. \citeyearpar{dillion} ask when models
can perform tasks previously assigned to human participants. The
corresponding survey-statistical question is which human-population
parameters are identified under a given protocol and information set,
which errors can be corrected, and what repetition gives an interval its
interpretation.

\citet{thompson2019} reviews how population surveys can combine data from sources such as censuses, administrative records, commercial databases, and nonprobability samples, and asks what general principles can be drawn from the different combinations. This review takes up that question for a source whose outcomes are generated by a model. When an LLM answers the questionnaire, the source of the respondent profiles and the mechanism generating their answers are distinct parts of the design. A survey can supply profiles, observed outcomes for validation, or residuals for a population estimator. Each role places different demands on the human data.

A probability survey connects observed units to its target through a
frame and inclusion probabilities. Stratified random sampling and
unequal-probability estimation provide the foundations of design-based
inference \citep{neyman, ht}. Coverage gaps, nonresponse, and
measurement error can still affect a probability survey. A human
benchmark should therefore state its population, weighting, and
precision \citep{groves, lumley}. Web-survey research shows that
convenient recruitment and large samples can coexist with selection
error \citep{bethlehem, baker}. Meng \citeyearpar{meng} connects
estimation error to the mechanism by which records enter a dataset. In a
generated survey, researchers may directly control profile composition,
while the response mechanism given a profile requires separate
validation. Repeated generation estimates the behavior of that mechanism
more precisely, leaving its difference from the target response
mechanism intact.

Comparative studies already show why validation depends on the estimand.
Argyle et al. \citeyearpar{argyle} reproduce some patterns of political
response by conditioning on characteristics drawn from human surveys.
Bisbee et al. \citeyearpar{bisbee} find that close aggregate means
coexist with differences in dispersion, subgroup results, and regression
coefficients. Xie et al. \citeyearpar{xie} extend the comparison to
cross-sectional and longitudinal benchmarks, exposing discrepancies in
life-event sequences and associations. Ahnert et al.
\citeyearpar{ahnert} and Park et al. \citeyearpar{park} show how
elicitation and profile information change performance. Prediction for a
new respondent from demographic records requires comparison with a
baseline that has the same information. Imputation for a respondent with
an existing interview requires held-out respondents or items and a
record of whether the interview already covered the target construct.

Identical marginal proportions can arise from different joint
distributions and subgroup contrasts. Population averages depend on both
the population shares and the conditional means, so discrepancies in the
two components can cancel. A similar distinction applies to survey
experiments. Ashokkumar et al. \citeyearpar{hewitt} find substantial
correlations between synthetic and human treatment estimates alongside
larger synthetic effect magnitudes. Cui et al. \citeyearpar{cui} show
why main effects and interactions need separate checks. Screening
candidate interventions depends on direction and ordering. Planning
sample sizes also requires effect magnitudes on the right scale and
information about their uncertainty. We therefore distinguish six levels
of evidence: individual responses, marginal distributions, joint
structure, subgroup contrasts, parameter estimation, and inferential
validity.

The way human outcomes enter the analysis determines how predictive
accuracy relates to inference. Model-assisted survey sampling combines
auxiliary predictions with residuals estimated from a probability sample
\citep{sarndal, deville}. Semisupervised mean estimation,
prediction-powered inference (PPI), and its efficiency extension
likewise distinguish the covariate sample used to average predictions
from the labeled sample used to correct residuals
\citep{zhangss, ppi, ppiplus}. Probability sampling identifies the
residual correction. Its precision depends on the covariance between the
predictions and observed responses. With a nonprobability sample,
carrying residual information to the population requires overlap and
assumptions about selection \citep{chendr, yangdr}. Separating
identification, efficiency, and coverage clarifies which empirical
checks matter after generated responses enter an estimator.

Existing reviews address neighboring questions. Sarstedt et al.
\citeyearpar{sarstedt} discuss silicon samples across stages of consumer
research, and Goyanes \citeyearpar{goyanes} reviews applications,
representation, and methodological challenges. Sen et al.
\citeyearpar{sen} trace measurement and representation error throughout
a simulated survey. D'Ecclesiis et al. \citeyearpar{decclesiis}
distinguish predictive, distributional, and substantive validity by use.
Feuerriegel et al. \citeyearpar{guide} provide a reporting checklist for
LLM studies in behavioral science. We connect these concerns to specific
estimators, showing how composition and conditional response
discrepancies enter target statistics and what is covered when
repetition is defined over tasks, respondents, or a survey design.

We organize the review around three statistical questions. What information enters the synthetic data, and which part of the joint distribution does it constrain? Which discrepancies persist when generated responses are used to estimate means, associations, subgroup contrasts, and experimental effects? Under which conditions can human survey data correct bias and support valid uncertainty statements? We use a common statistical framework to compare methods, worked examples to show when their conclusions differ, and a paired reanalysis of two public benchmarks to assess the frequency and magnitude of effect amplification.

Figure~\ref{fig:1} summarizes the workflow. Sections 2--4 define the statistical
objects, generation choices, and adjustments. Section 5 develops the
bias, coverage, and correction results. Section 6 synthesizes the
evidence and presents the paired reanalysis. Section 7 discusses
validation designs and practical choices. Section 8 concludes.

\begingroup\setstretch{1}
\begin{figure}
\centering
\includegraphics[width=0.94\textwidth,height=\textheight]{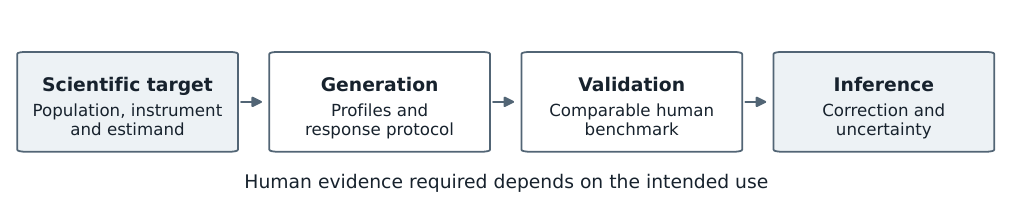}
\caption{Research workflow. The estimand determines the benchmark and
inferential requirements.}
\label{fig:1}
\end{figure}
\endgroup

\section{Scope and statistical
foundations}\label{scope-and-statistical-foundations}

\subsection{Objects and conceptual
boundaries}\label{objects-and-conceptual-boundaries}

An \emph{LLM-generated survey response} is an answer elicited from a
language model using a survey item, its response format, and a specified
context. An \emph{LLM-generated survey sample} is a collection of
records that links these answers to actual or constructed profiles under
a reproducible protocol. We use \emph{synthetic respondent} for the unit
being simulated and \emph{silicon sampling} for the practice of
eliciting such responses for a set of profiles. A directly elicited
vector of answer shares is an aggregate prediction. It becomes a set of
synthetic records only after a sampling rule is specified. If the model
produces an answer distribution for each item separately, the implied
record file depends on how those distributions are coupled.

Table~\ref{tab:1} identifies the statistical connection with neighboring methods.
In disclosure synthesis, inference starts from collected microdata and
depends on the synthesis design and associated variance rules
\citep{reiter}. In model-based imputation, the frame of observed cases
and the missingness mechanism define the problem. Generating a response
for an actual respondent with an unobserved outcome can be treated as
imputation, provided the analysis accounts for the missing-data
mechanism and imputation uncertainty. An entirely generated panel
introduces the additional problem of specifying the distribution of
profiles and their transitions.

\begingroup\setstretch{1}
\begin{longtable}[]{@{}
  >{\raggedright\arraybackslash}p{(\columnwidth - 4\tabcolsep) * \real{0.3333}}
  >{\raggedright\arraybackslash}p{(\columnwidth - 4\tabcolsep) * \real{0.3333}}
  >{\raggedright\arraybackslash}p{(\columnwidth - 4\tabcolsep) * \real{0.3333}}@{}}
\caption{Boundaries of survey response synthesis}\label{tab:1}\\
\toprule\noalign{}
\begin{minipage}[b]{\linewidth}\raggedright
Task
\end{minipage} & \begin{minipage}[b]{\linewidth}\raggedright
Information entering the procedure
\end{minipage} & \begin{minipage}[b]{\linewidth}\raggedright
Output and target
\end{minipage} \\
\midrule\noalign{}
\endfirsthead
\toprule\noalign{}
\begin{minipage}[b]{\linewidth}\raggedright
Task
\end{minipage} & \begin{minipage}[b]{\linewidth}\raggedright
Information entering the procedure
\end{minipage} & \begin{minipage}[b]{\linewidth}\raggedright
Output and target
\end{minipage} \\
\midrule\noalign{}
\endhead
\bottomrule\noalign{}
\endlastfoot
Synthetic respondents & Profiles, instrument, model and elicitation
protocol & Synthetic answers or records for an explicitly defined
reference population \\
Aggregate response prediction & Item, options, target group and optional
benchmark information & Group-level answer shares or another forecast \\
Missing-response imputation & Observed units, partial responses and
missingness indicators & Completed observed cases with uncertainty tied
to the imputation procedure \\
Disclosure synthesis & Collected confidential records and a release
model & Released data with synthesis-specific utility and disclosure
properties \\
Statistical simulation & A prespecified or estimated data-generating
process & Replications for checking an estimator or testing a
mechanism \\
Agent-based simulation & Agent states, transition and interaction rules
& Trajectories or system outcomes, including survey responses when
elicited \\
\end{longtable}
\endgroup

\subsection{Target, reference survey, and simulation
law}\label{target-reference-survey-and-simulation-law}

Let \(P\) be the joint distribution of covariates \(X\) and human
responses \(Y\) under a stated population, time, and questionnaire
protocol. The observed survey has a law \(P_s\) determined by coverage,
sampling, response, and measurement. Survey weights and adjustment
models link the observed data to \(P\) under their own conditions. A
comparison against an unweighted set of complete cases targets that set,
or a model for it. A comparison against weighted estimates can address
population quantities when the survey adjustments support them.
Benchmark uncertainty matters in either case. Coverage, nonresponse,
measurement, and processing errors can enter the human benchmark.
Comparing the same instrument, retaining survey weights, and reporting
benchmark precision help distinguish survey errors from generation
errors \citep{groves}.

Write \(Q(dx,dy)=Q_X(dx)Q(dy\mid x)\) for the simulation law. The
protocol fixes the model release, instructions, question wording, option
labels and order, language, decoding, parsing, and treatment of
refusals. It also fixes any examples, retrieval corpus, or adaptive
changes. If the prompt uses additional information \(Z\), the
conditional law averages over the distribution of \(Z\) supplied at a
given \(X\). A generated profile with a plausible demographic label
therefore conveys a different information set from a linked transcript
recorded from a real person. Figure~\ref{fig:2} identifies the information paths
that connect human outcomes to the model and to the final estimator.

\begingroup\setstretch{1}
\begin{figure}
\centering
\includegraphics[width=0.95\textwidth,height=\textheight]{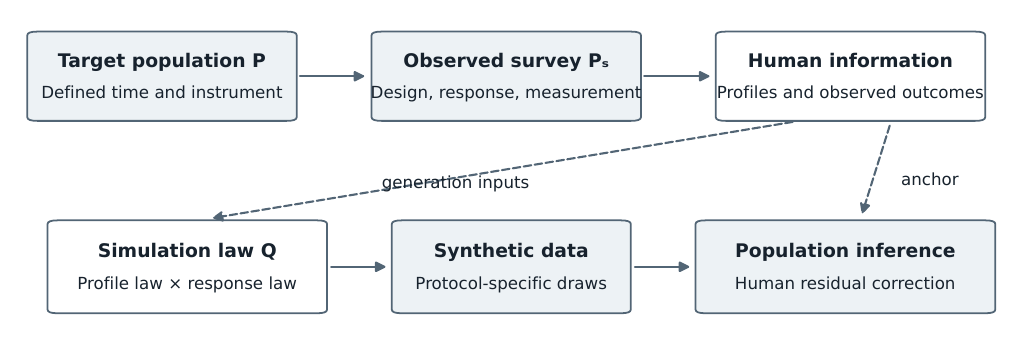}
\caption{The population, observed survey, and simulation law. Human
outcome data can enter training, independent assessment, and estimation
through distinct paths.}
\label{fig:2}
\end{figure}
\endgroup

\section{A taxonomy of generated survey
samples}\label{a-taxonomy-of-generated-survey-samples}

The axes in Figure~\ref{fig:3} describe separate choices. A study can reuse survey
profiles, elicit a battery jointly, fine-tune on earlier waves, and
apply post-generation correction to a single target mean. These choices
constrain different properties. Profile reuse fixes the empirical
mixture of input profiles, conditional on the restrictions and weights
applied. Joint elicitation permits within-record dependence. Fine-tuning
can change the response law. An estimating-equation correction changes
the estimator for a specified target.

\begingroup\setstretch{1}
\begin{figure}
\centering
\includegraphics[width=0.96\textwidth,height=\textheight]{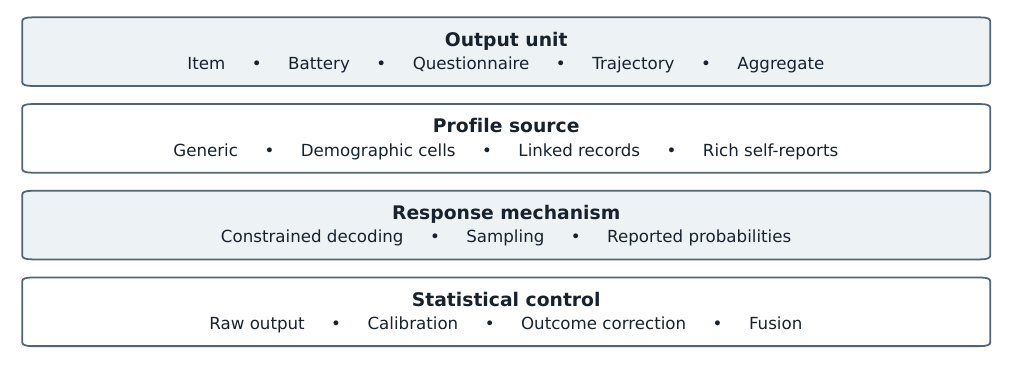}
\caption{Four design axes. Each protocol chooses a level on each axis,
with combinations across axes permitted.}
\label{fig:3}
\end{figure}
\endgroup

\begingroup\setstretch{1}
\begin{longtable}[]{@{}
  >{\raggedright\arraybackslash}p{(\columnwidth - 4\tabcolsep) * \real{0.3333}}
  >{\raggedright\arraybackslash}p{(\columnwidth - 4\tabcolsep) * \real{0.3333}}
  >{\raggedright\arraybackslash}p{(\columnwidth - 4\tabcolsep) * \real{0.3333}}@{}}
\caption{Method choices and the assumptions that govern their use}\label{tab:2}\\
\toprule\noalign{}
\begin{minipage}[b]{\linewidth}\raggedright
Choice
\end{minipage} & \begin{minipage}[b]{\linewidth}\raggedright
Statistical role and strength
\end{minipage} & \begin{minipage}[b]{\linewidth}\raggedright
Main dependency and use
\end{minipage} \\
\midrule\noalign{}
\endfirsthead
\toprule\noalign{}
\begin{minipage}[b]{\linewidth}\raggedright
Choice
\end{minipage} & \begin{minipage}[b]{\linewidth}\raggedright
Statistical role and strength
\end{minipage} & \begin{minipage}[b]{\linewidth}\raggedright
Main dependency and use
\end{minipage} \\
\midrule\noalign{}
\endhead
\bottomrule\noalign{}
\endlastfoot
Fixed, matched, or externally sampled profiles & Controls the mix of
simulated units & Population frame, weights, support, and accessible
covariates govern population interpretation \\
Generated profiles & Produces records where frames are sparse & Both
profile composition and conditional responses require validation \\
Prompting or constrained decoding & Changes response elicitation at low
setup cost & Option order, parsing, refusal, and model version define
the measured law \\
Retrieval, examples, or fine-tuning & Introduces observed human
information & Gain depends on the proximity of source and target,
information leakage, and deployment availability \\
Joint or sequential generation & Allows cross-item and cross-wave
dependence & Dependence and order effects require direct validation
against linked records \\
Calibration or matching & Aligns measured auxiliary features & Retained
support and relevant balance functions determine remaining composition
error \\
Human-anchored correction & Uses observed outcomes to estimate
target-specific discrepancies & A valid human design or transport
assumption identifies the correction \\
\end{longtable}
\endgroup

The \textbf{unit} axis ranges from a single item through a battery, a
full questionnaire, and an interactive respondent trajectory. It also
includes direct distribution or effect forecasts, which produce
aggregate objects. The \textbf{profile} axis ranges from a generic
persona through demographic cells and sampled microdata profiles to
linked responses, interviews, and time-specific context. The
\textbf{response} axis covers label probabilities, sampled outputs,
verbalized distributions, and joint or sequential protocols, with
prompting, retrieval, and fine-tuning as mechanisms for conditioning or
adaptation. The \textbf{statistical control} axis covers raw output,
design or post-stratification weights, balancing, outcome calibration,
and correction of an estimator. Assigning a study to all four axes
clarifies where observed information enters. Record the information
available at deployment, the data available during development, and the
human data set aside for evaluation. A single model and prompt can
implement different tasks under different information restrictions.

\section{Methodological developments}\label{methodological-developments}

\subsection{From personas to instrument
design}\label{from-personas-to-instrument-design}

Matched backstories standardize the composition of a comparison and make
it possible to evaluate conditional performance \citep{argyle, bisbee}.
The evaluator should record which attributes are available at
deployment. Party identification or prior attitudes can supply
considerable signal when forecasting a political item. Their use changes
the substantive task if the proposed application begins with
demographics alone. Likewise, an interview containing an answer
equivalent to a test item changes an apparent prediction task into
retrieval. Evaluation should hold out respondents, constructs, and
survey waves according to the intended deployment. Within-person
prediction from an observed history is appropriate for a panel
imputation scenario, where that history is part of the information set.

Rich profiles improve some outcomes through information about
individuals rather than additional draws from a fixed persona. Park et
al. \citeyearpar{park} compare demographic profiles, self-authored
personas, questionnaires, and interview transcripts for 1,052 US adults.
In their November 2026 revision, combining interview and survey
information raises held-out GSS item accuracy from 60.4\% with
demographics to 72.8\%, against 79.5\% human test--retest agreement. The
same information improves several measures of distortion, while
prediction of incentivized economic choices shows weaker gains. The
associated collection cost belongs in a comparison with direct survey
measurement. Suh et al. \citeyearpar{suh} pursue a different objective:
supervised learning of subgroup response distributions from SubPOP.
Their improvement in distributional prediction uses observed survey
shares in training and should be assessed against baselines with the
same source data. Jiang et al. \citeyearpar{jiang} illustrate why a
human-history-only model also provides a relevant comparison. Profiles,
historical items, and directly observed outcomes supply different
information and incur separate costs.

\subsection{Elicitation and response
structure}\label{elicitation-and-response-structure}

The elicitation protocol regulates the relationship between model tokens
and questionnaire answers. First-token probabilities aggregate over a
limited labeling scheme. Constrained generation changes the set of
admissible outputs. An open response can be mapped to a category through
a separate classification step. Directly elicited probabilities are
numeric reports from a language model. Meister et al.
\citeyearpar{meister} show that reported distributions can align more
closely with human distributions than distributions produced by repeated
response simulation. Ahnert et al. \citeyearpar{ahnert} compare eight
response generation methods and find that restricted generation performs
best overall in their closed-ended political tasks. These findings
concern different comparison units and objectives. An accurate
probability vector can support an item-level forecast while leaving
within-person dependence unspecified.

Instrument perturbations identify parts of the simulation law that a
single benchmark match conceals. Dominguez-Olmedo et al.
\citeyearpar{dominguez} document choice-label and order effects on
American Community Survey items. Under randomized answer order, many
tested model distributions moved toward uniformity. Tjuatja et al.
\citeyearpar{tjuatja} find that changing wording or design does not
reliably reproduce the corresponding human response effect across their
nine tested models. These studies test model sensitivity to measurement
design. When human measurement is order-sensitive, the direct comparison
is between human and synthetic effects under the same randomization.
When the human protocol fixes the order, averaging over arbitrary
permutations defines a different target for the simulation.

When models report answer probabilities, use a strictly proper scoring
rule against baselines with the same inputs \citep{gneiting}. Let
\(v=(v_1,\ldots,v_J)\) be the reported probability vector and
\(p_j(X)=P(Y=j\mid X)\) the human conditional probability for category
\(j\). Taking the conditional expectation of the quadratic probability
loss gives

\begin{equation}
\label{eq:1}
E_P\!\left[\sum_{j=1}^J (v_j-\mathbf 1\{Y=j\})^2\mid X\right]
=\sum_{j=1}^J (v_j-p_j)^2+1-\sum_{j=1}^J p_j^2.
\end{equation}

The last two terms are independent of the forecast, so the expected loss
is minimized at \(v=p\). Compare scores on the same held-out sample and
apply survey weights where appropriate. This comparison assesses
probability forecasts for a single item. Assessing dependence between
items requires a joint outcome.

For multiple items, generating each response independently given a
profile can reproduce every conditional marginal while setting
within-profile item covariance to zero. A joint prompt or a sequential
protocol with respondent-level memory makes such covariance possible. It
also introduces dependencies on item order, context length, and prior
synthetic responses. Assessment should include joint cell frequencies,
within-profile covariance, scales, skip patterns, and missingness. A
generated reason is useful for instrument review when analysts can trace
it to a concrete problem with wording. Its frequency remains a statistic
of the protocol until verified against human data.

\subsection{Calibration and allocation of human
data}\label{calibration-and-allocation-of-human-data}

Margin calibration, joint-cell calibration, outcome correction, and
fusion solve distinct statistical problems. Raking can align demographic
margins while retaining errors in interactions. Joint-cell weighting
targets specified cross-classifications when their population counts and
support are available. Outcome correction uses observed answers to
estimate the remaining difference for a particular statistic. It can
improve a mean without changing any record in the generated file. This
distinction is especially important when synthetic records are released
for secondary analysis, where a correction calculated for one outcome
supplies no general guarantee for other estimands.

Krsteski et al. \citeyearpar{krsteski} compare prompting, fine-tuning,
prediction-powered inference (PPI), and design-based supervised learning
(DSL) for selected mean targets from two human surveys. Their comparison
uses the benchmark dataset means as targets and supports reserving
gold-standard labels for correction under the budgets they study. The
tuned correction reaches an estimated effective sample size gain of
about 14\% in one dietary-intake setting. Unit-weight correction gives
negative effective sample size gains in their reported comparison. A
fair comparison charges human labels used for prompt development,
fine-tuning, tuning the correction, and evaluation to their respective
roles. At a fixed acquisition budget, additional training labels are
valuable if they improve predictions enough to reduce residual variance.
Additional correction labels directly improve estimation of that
residual. Their relative value depends on the target statistic, the
stability of the response model, and the size and quality of the target
human sample.

\section{Statistical properties and
inference}\label{statistical-properties-and-inference}

\subsection{Target means, bias decomposition, and
calibration}\label{target-means-bias-decomposition-and-calibration}

Let \(P\) be the joint law in the target population and \(Q\) the joint
law under a fixed generation protocol. For an integrable function
\(h(X,Y)\), the estimand is \(\theta=E_Ph\). Write
\(m_R(x)=E_R[h(X,Y)\mid X=x]\) for \(R=P,Q\). Assume the conditional
means are defined on the supports used in the comparison. By iterated
expectation, \(E_Rh=E_{R_X}m_R\). Adding and subtracting \(E_{P_X}m_Q\)
yields

\begin{equation}
\label{eq:2}
E_Qh-\theta
=(E_{Q_X}-E_{P_X})m_Q+E_{P_X}(m_Q-m_P).
\end{equation}

Denote the two terms by \(B_{\mathrm{s}}\) and \(B_{\mathrm{c}}\),
respectively. The first changes the profile distribution while holding
the generated conditional mean fixed. The second compares response
mechanisms under the target profile distribution. Adding and subtracting
\(E_{Q_X}m_P\) instead gives \((E_{Q_X}-E_{P_X})m_P\) and
\(E_{Q_X}(m_Q-m_P)\). The components depend on the order of replacement;
report which distribution serves as the reference.

For \(G\) disjoint groups, let \(p_g=P(X=g)\) and \(q_g=Q(X=g)\), and
let \(\mu_g\) and \(\nu_g\) be the human and generated conditional
means. Equation~\eqref{eq:2} becomes

\begin{equation}
\label{eq:3}
B_{\mathrm{s}}=\sum_{g=1}^{G}(q_g-p_g)\nu_g,
\qquad B_{\mathrm{c}}=\sum_{g=1}^{G}p_g(\nu_g-\mu_g).
\end{equation}

For example, suppose the human shares are \((0.50,0.50)\) with
conditional proportions \((0.30,0.70)\), and the generated shares are
\((0.70,0.30)\) with conditional proportions \((0.40,0.80)\). The target
proportion is \(0.50\times0.30+0.50\times0.70=0.50\), and the generated
proportion is \(0.70\times0.40+0.30\times0.80=0.52\). The composition
component is \(0.20\times0.40-0.20\times0.80=-0.08\). The conditional
response component is \(0.50\times0.10+0.50\times0.10=0.10\). The total
bias is \(0.02\). After calibration to the correct shares, the generated
proportion is \(0.50\times0.40+0.50\times0.80=0.60\). The smaller
initial bias arose from cancellation, which the separate components make
visible.

Knowing \(P_X\) and \(Q\) still allows two human joint distributions to
have the same profile distribution and different conditional means on a
set with positive probability. The observed inputs are then identical
while the target means differ. Observed outcomes or restrictions on
human responses determine what can be identified \citep{manski}.

Let \(w(X)\) be a nonnegative calibration weight with \(E_Qw=1\), and
denote the weighted law by \(Q_w\). For profiles with positive weight,
weighting by \(X\) alone preserves \(Q(Y\in A\mid X=x)\) for any
response event \(A\). Let \(b(X)\) be a vector of balance functions and
impose \(E_{Q_w}b=E_Pb\). Write \(m_Q(X)=b(X)^\top\gamma+r(X)\).
Substitution into~\eqref{eq:2} cancels \(\gamma^\top(E_{Q_w}b-E_Pb)\), leaving

\begin{equation}
\label{eq:4}
E_{Q_w}h-\theta
=(E_{Q_{w,X}}-E_{P_X})r+E_{P_X}(m_Q-m_P).
\end{equation}

The balance constraints remove the part of composition bias explained by
\(b\). If an age-by-sex interaction matters to \(m_Q\) while only
separate margins are balanced, the remainder \(r\) can still cause
composition bias. Joint-cell calibration can reduce that component when
cell totals and adequate support are available \citep{deville, sarndal}.

If \(Q_{w,X}\) is absolutely continuous with respect to \(P_X\), let
\(\rho=dQ_{w,X}/dP_X\). When \(r\) and \(\rho-1\) are square-integrable
under \(P_X\), Cauchy--Schwarz gives

\begin{equation}
\label{eq:5}
\left|(E_{Q_{w,X}}-E_{P_X})r\right|
=\left|E_{P_X}[(\rho-1)r]\right|
\le \{E_{P_X}(\rho-1)^2\}^{1/2}\{E_{P_X}r^2\}^{1/2}.
\end{equation}

The bound expresses the separate roles of profile imbalance and
approximation error in the balance model. Weight dispersion, limited
support, and unmodeled interactions call for separate checks.

\subsection{Dispersion, associations, and experimental
effects}\label{dispersion-associations-and-experimental-effects}

For \(h=Y\), write \(m_R^Y(X)=E_R(Y\mid X)\). Decompose \(Y-E_RY\) into
\(Y-m_R^Y(X)\) and \(m_R^Y(X)-E_RY\). The first term has conditional
mean zero, so the expected cross-product vanishes. Squaring and taking
expectations gives

\begin{equation}
\label{eq:6}
\operatorname{Var}_R(Y)
= E_R\{\operatorname{Var}_R(Y\mid X)\}
+\operatorname{Var}_R\{m_R^Y(X)\}.
\end{equation}

For two responses, let \(m_{Rj}(X)=E_R(Y_j\mid X)\) and expand the
centered product in the same way.

\begin{equation}
\label{eq:7}
\operatorname{Cov}_R(Y_1,Y_2)
=E_R\{\operatorname{Cov}_R(Y_1,Y_2\mid X)\}
+\operatorname{Cov}_R\{m_{R1}(X),m_{R2}(X)\}.
\end{equation}

Independent item generation given a profile sets the first term to zero.
Between-profile differences in conditional means can still induce
marginal association. Scale reliability, battery dependence, and
longitudinal transitions require checking both sources of association.

For a linear projection with an intercept and finite second moments, let
\(\Sigma_R=\operatorname{Var}_R(X)\) and
\(c_R=\operatorname{Cov}_R(X,Y)\). The squared-error normal equations
are \(\Sigma_R\beta_R=c_R\). Subtract \(\Sigma_Q\beta_P\) and use
\(c_P=\Sigma_P\beta_P\). With nonsingular covariance matrices, this
gives

\begin{equation}
\label{eq:8}
\beta_Q-\beta_P
=\Sigma_Q^{-1}\{(c_Q-c_P)-(\Sigma_Q-\Sigma_P)\beta_P\}.
\end{equation}

Changes in the response--covariate cross-moment and the covariate
covariance both affect the slopes. A nearly singular \(\Sigma_Q\) can
magnify small moment discrepancies. For example, suppose two binary
variables each have marginal probability \(0.5\). If \(Y=X\) in the
human population, then \(E_P(XY)=0.5\) and
\(\operatorname{Cov}_P(X,Y)=0.5-0.25=0.25\). If they are independent
under generation, the covariance is \(0.25-0.25=0\). Since
\(\operatorname{Var}(X)=0.25\) in both laws, the slopes are \(1\) and
\(0\).

For two groups \(g=0,1\), define \(b_g=\nu_g-\mu_g\). The error in the
difference between group means is

\begin{equation}
\label{eq:9}
(\nu_1-\nu_0)-(\mu_1-\mu_0)=b_1-b_0.
\end{equation}

If the groups have equal population shares, the human group means are
\(0.30\) and \(0.70\), and both generated means are \(0.50\), the
overall means agree at \(0.50\). The contrast has error
\((-0.20)-(0.20)=-0.40\). Comparisons between groups require the size
and direction of both group errors.

In a survey experiment, let \(Y(a)\) be the potential response under
treatment \(a\in\{0,1\}\). Random assignment, consistency, and no
interference underlie interpretation of the human contrast as an average
treatment effect in the relevant experimental setting
\citep{imbens, linreg}. Define \(\mu_a^R(x)=E_R[Y(a)\mid X=x]\),
\(\tau_R(x)=\mu_1^R(x)-\mu_0^R(x)\), and
\(\delta_a(x)=\mu_a^Q(x)-\mu_a^P(x)\). When each experiment is averaged
under its own profile distribution, add and subtract \(E_{P_X}\tau_Q\)
and use \(\tau_Q-\tau_P=\delta_1-\delta_0\).

\begin{equation}
\label{eq:10}
E_{Q_X}\tau_Q-E_{P_X}\tau_P
=(E_{Q_X}-E_{P_X})\tau_Q+E_{P_X}(\delta_1-\delta_0).
\end{equation}

Standardization to a common profile distribution removes the first term.
The remaining error is the difference between conditional response
biases in the two arms. For example, doubling every human effect
preserves its ordering and doubles the magnitude of every nonzero
effect. Section 6.3 examines frequency and magnitude separately.

\subsection{Generated sample size and interval
coverage}\label{generated-sample-size-and-interval-coverage}

Let \(H_1,\ldots,H_M\) be independent draws of \(h(X,Y)\) under fixed
\(Q\), and write \(\mu_Q=E_Qh\) and
\(\sigma_Q^2=\operatorname{Var}_Q(h)\). Decompose \(\bar H_M-\theta\)
into \(\bar H_M-\mu_Q\) and \(\mu_Q-\theta\). The former has mean zero,
eliminating the cross-term \(2(\mu_Q-\theta)E_Q(\bar H_M-\mu_Q)\).
Independence gives

\begin{equation}
\label{eq:11}
E_Q(\bar H_M-\theta)^2
=\frac{\sigma_Q^2}{M}+(\mu_Q-\theta)^2.
\end{equation}

Increasing \(M\) lowers the first term. If \(b=\mu_Q-\theta\ne0\) and an
interval centered at \(\bar H_M\) has half-width \(O_p(M^{-1/2})\), take
\(\varepsilon=|b|/3\). When the center lies within \(\varepsilon\) of
\(\mu_Q\) and the half-width is below \(\varepsilon\), the interval
excludes \(\theta\). Both events have probability tending to one. For a
sequence of laws \(Q_M\) with standard deviations tending to
\(\sigma_Q>0\), suppose \(\sqrt M b_M/\sigma_Q\to c\) and the centered
statistic satisfies a central limit theorem. Then the symmetric interval
using \(\sigma_Q\) has limiting coverage

\begin{equation}
\label{eq:12}
\Pr_{Q_M}(\theta\in I_M)
\longrightarrow \Pr(-z\le Z+c\le z)
=\Phi(z-c)-\Phi(-z-c),\qquad Z\sim N(0,1).
\end{equation}

At \(c=0\) the limit equals the nominal level. This result identifies
the scale on which bias must be compared with standard error.

For independent Bernoulli generation, write \(q\) for the generation
probability, \(p\) for the human target, and \(S_M\) for the number of
successes. The Wilson interval inverts the score test
\citep{wilson, brown}. It contains \(p\) exactly when
\(|S_M-Mp|\le z\sqrt{Mp(1-p)}\). Set
\(L_M=\max\{0,\lceil Mp-z\sqrt{Mp(1-p)}\rceil\}\) and
\(U_M=\min\{M,\lfloor Mp+z\sqrt{Mp(1-p)}\rfloor\}\). The exact coverage
is

\begin{equation}
\label{eq:13}
C(M,p,q)=\Pr_q(L_M\le S_M\le U_M)
=\sum_{s=L_M}^{U_M}\binom{M}{s}q^s(1-q)^{M-s}.
\end{equation}

At nominal coverage \(0.95\), \(p=0.45\), and \(q=0.50\), the continuous
endpoints at \(M=100\) are approximately \(35.25\) and \(54.75\), giving
accepted counts \(36\) through \(54\). At \(M=1000\), the accepted
counts are \(420\) through \(480\). The sums in~\eqref{eq:13} give coverage of
\textbf{81.4\%} and \textbf{10.9\%}, respectively. Figure~\ref{fig:4} uses the
same calculation.

\begingroup\setstretch{1}
\begin{figure}
\centering
\includegraphics[width=0.91\textwidth,height=\textheight]{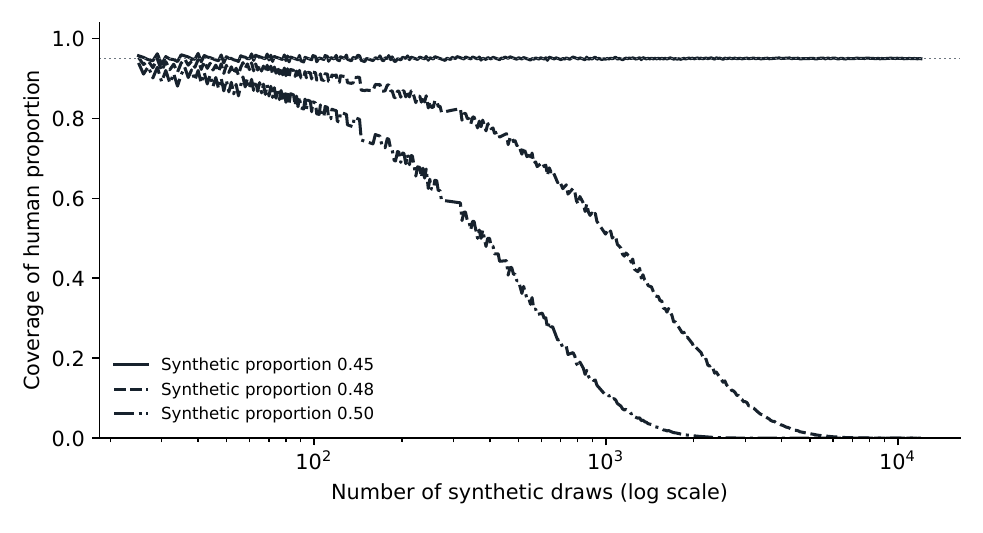}
\caption{Exact coverage of nominal 95\% Wilson intervals for a human
target proportion of 0.45 under independent Bernoulli generation.}
\label{fig:4}
\end{figure}
\endgroup

Repeated responses to the same profile share a profile even when draws
are independent conditional on it. Let \(K\) independent profiles each
receive \(r\) draws. Write \(\bar Y_k\) for the mean for profile \(k\),
\(A=\operatorname{Var}_{Q_X}\{m_Q^Y(X)\}\), and
\(B=E_{Q_X}\{\operatorname{Var}_Q(Y\mid X)\}\). The conditional mean of
\(\bar Y_k\) is \(m_Q^Y(X_k)\), and its conditional variance is
\(\operatorname{Var}(Y\mid X_k)/r\). Applying~\eqref{eq:6} and independence
across profiles gives

\begin{equation}
\label{eq:14}
\operatorname{Var}\left(K^{-1}\sum_{k=1}^K\bar Y_k\right)
=\frac{1}{K}\left(A+\frac{B}{r}\right).
\end{equation}

Additional draws for a fixed profile reduce \(B/r\). Let a new profile
cost \(c_x>0\), each additional generation cost \(c_r>0\), and the
budget be \(C\). Substituting \(K=C/(c_x+c_rr)\) into~\eqref{eq:14} leads to the
continuous optimum, for \(A,B>0\),

\begin{equation}
\label{eq:15}
\frac{d}{dr}\frac{(A+B/r)(c_x+c_rr)}{C}
=\frac{A c_r-Bc_x/r^2}{C},
\qquad r^*=\sqrt{\frac{Bc_x}{A c_r}}.
\end{equation}

Use the continuous optimum to guide comparison of feasible integer
allocations. Large between-profile variation favors collecting more
distinct profiles. High within-profile variation and expensive profiles
favor repeated draws. Learning and version changes add separate sources
of variability when draws share a fitted model.

\subsection{Residual correction with a probability
sample}\label{residual-correction-with-a-probability-sample}

Consider a finite population of size \(N\) with human outcomes \(Y_i\)
and fixed auxiliary predictions \(f_i\) available throughout the frame.
Their population mean \(\bar f_U\) is known. Let \(I_i\) indicate
inclusion in a probability sample, with \(\pi_i=E_d(I_i)>0\), where
\(d\) indicates the sampling design. Put
\(e_i(\lambda)=Y_i-\lambda f_i\). The model-assisted difference
estimator with fixed \(\lambda\) is \citep{ht, sarndal}

\begin{equation}
\label{eq:16}
\widehat\theta_\lambda
=\lambda\bar f_U+\frac{1}{N}\sum_{i=1}^N\frac{I_i e_i(\lambda)}{\pi_i}.
\end{equation}

The first term is the average prediction and the second estimates its
residual. Since \(E_d(I_i/\pi_i)=1\),

\begin{equation}
\label{eq:17}
E_d(\widehat\theta_\lambda)
=\lambda\bar f_U+\frac1N\sum_{i=1}^N(Y_i-\lambda f_i)
=\bar Y_U.
\end{equation}

Thus the estimator is design-unbiased with fixed predictions, fixed
weight, and a known population prediction mean. With
\(\pi_{ij}=E_d(I_iI_j)\) and \(\pi_{ii}=\pi_i\), the covariance of the
inclusion indicators is \(\pi_{ij}-\pi_i\pi_j\). This yields

\begin{equation}
\label{eq:18}
\operatorname{Var}_d(\widehat\theta_\lambda)
=\frac1{N^2}\sum_{i=1}^N\sum_{j=1}^N
(\pi_{ij}-\pi_i\pi_j)\frac{e_i(\lambda)e_j(\lambda)}{\pi_i\pi_j}.
\end{equation}

The variance can be estimated from the design when second-order
inclusion probabilities are available and positive. Complex surveys
require compatible linearization or replicate weights, with the relevant
correction repeated for each replicate \citep{raowu, lumley}. If the
predictor or coefficient is estimated with the same outcomes, account
for learning in the expectation and variance. Cross-fitting can separate
prediction training from residual evaluation, subject to the conditions
of the sampling design \citep{dml}.

Under simple random sampling without replacement, let \(n\) be the
sample size and let \(S_Y^2\), \(S_f^2\), and \(S_{Yf}\) be the
finite-population variances and covariance. Equation~\eqref{eq:18} reduces to

\begin{equation}
\label{eq:19}
\operatorname{Var}_d(\widehat\theta_\lambda)
=\frac{1-n/N}{n}\bigl(S_Y^2-2\lambda S_{Yf}+\lambda^2S_f^2\bigr).
\end{equation}

If \(S_f^2>0\), differentiation gives \(\lambda^*=S_{Yf}/S_f^2\).
Compared with the direct estimate at \(\lambda=0\), the optimized
residual variance decreases by \(S_{Yf}^2/S_f^2\). With \(\lambda=1\)
and \(n<N\), precision improves exactly when \(2S_{Yf}>S_f^2\). Observed
residuals supply the identification. Prediction dispersion and its
covariance with the outcome govern the precision gain.

If the prediction mean is unknown, a separate sample of \(M_X\) observed
covariate profiles can estimate it. Take the labeled and covariate-only
samples to be independent and each drawn i.i.d. from the same target
law, with fixed \(f\) and \(\lambda\) \citep{zhangss, ppi, ppiplus}. Let
\(e_i(\lambda)=Y_i-\lambda f(X_i)\) in the labeled sample. The
prediction-powered estimator is

\begin{equation}
\label{eq:20}
\widehat\theta^{\mathrm{PPI}}_\lambda
=\frac{1}{n}\sum_{i=1}^{n}e_i(\lambda)
+\frac{\lambda}{M_X}\sum_{j=1}^{M_X}f(X'_j).
\end{equation}

The expectations of the two terms add to \(E_PY\). The samples are
independent, and expanding the residual variance gives

\begin{equation}
\label{eq:21}
\operatorname{Var}(\widehat\theta^{\mathrm{PPI}}_\lambda)
=\frac{\sigma_Y^2-2\lambda\sigma_{Yf}+\lambda^2\sigma_f^2}{n}
+\frac{\lambda^2\sigma_f^2}{M_X}.
\end{equation}

Here \(\sigma_{Yf}=\operatorname{Cov}(Y,f)\) and
\(\sigma_f^2=\operatorname{Var}(f)\). Differentiating in \(\lambda\)
gives \(-2\sigma_{Yf}/n+2\lambda\sigma_f^2(1/n+1/M_X)=0\). For
\(\sigma_f^2>0\), the solution is

\begin{equation}
\label{eq:22}
\lambda^*=
\frac{\sigma_{Yf}}{\sigma_f^2(1+n/M_X)}.
\end{equation}

A smaller covariate sample increases uncertainty in the prediction mean
and shrinks the optimal coefficient in magnitude. Additional generations
for the same profiles do not increase \(M_X\). For \(n=100\),
\(M_X=900\), \(\sigma_Y^2=0.25\), \(\sigma_f^2=0.09\), and
\(\sigma_{Yf}=0.09\), the optimum is \(0.90\). Equation~\eqref{eq:21} gives
\((0.25-0.162+0.0729)/100+0.0729/900=0.00169\). Direct estimation gives
\(0.25/100=0.00250\), so the reduction is \(32.4\%\). The equivalent
direct-estimation sample size is \(0.25/0.00169\approx148\).
Design-based supervised learning likewise uses observed residuals to
correct predictions, with a labeling design aligned to the inferential
target \citep{dsl}.

\subsection{Nonprobability benchmarks and combining
estimators}\label{nonprobability-benchmarks-and-combining-estimators}

Let \(A=1\) indicate a unit with a human outcome and
\(\pi_0(X)=\Pr(A=1\mid X)\). For this discussion, \(P\) also denotes the
law of \((X,Y,A)\). Suppose \(E(Y\mid X,A=1)=m_P^Y(X)\), there is
adequate overlap, and \(P_X\) is known or estimable from the target
population. Combine an outcome model \(f(X)\) and an inclusion model
\(\pi(X)>0\) \citep{littlerubin, chendr, yangdr}. The conditional
expectation of the weighted residual is
\(\pi_0(X)\{m_P^Y(X)-f(X)\}/\pi(X)\). Therefore the population bias is

\begin{equation}
\label{eq:23}
E_P\left[f(X)+\frac{A\{Y-f(X)\}}{\pi(X)}\right]-E_PY
=E_{P_X}\left[\left(1-\frac{\pi_0(X)}{\pi(X)}\right)
                    \{f(X)-m_P^Y(X)\}\right].
\end{equation}

Under these identification conditions, either a correct inclusion model
or a correct outcome model is sufficient for zero bias. The
probabilities used to assign profiles to a generation protocol describe
its simulation design. Applying~\eqref{eq:23} to a human population requires
human outcomes and an account of their selection into the benchmark.

When conditional mean ignorability is uncertain, fix the target profile
mix and perform a sensitivity analysis. Let \(R=Y-f(X)\),
\(d(x)=E(R\mid X=x,A=1)-E_P(R\mid X=x)\), and assume \(|d(x)|\le b(x)\).
Let \(\theta_{\mathrm{corr}}=E_{P_X}[f(X)+E(R\mid X,A=1)]\). The
triangle inequality gives

\begin{equation}
\label{eq:24}
|\theta_{\mathrm{corr}}-E_PY|
=|E_{P_X}d(X)|\le E_{P_X}b(X).
\end{equation}

For a subgroup contrast, add the two group-specific bounds. This
expresses an untestable transport assumption as a range of potential
error \citep{manski}.

Consider an unbiased probability-sample estimator \(\widehat\theta_P\)
with variance \(V_P\), and a generated estimator \(\widehat\theta_G\)
with bias \(b\) and variance \(V_G\). Let \(C_{PG}\) denote their
covariance. For fixed \(w\in[0,1]\), define
\(\widehat\theta_w=(1-w)\widehat\theta_P+w\widehat\theta_G\). Its bias
is \(wb\), and its risk is

\begin{equation}
\label{eq:25}
\operatorname{MSE}(\widehat\theta_w)
=(1-w)^2V_P+w^2(V_G+b^2)+2w(1-w)C_{PG}.
\end{equation}

Expanding the quadratic gives
\(V_P+2w(C_{PG}-V_P)+w^2(V_P+V_G+b^2-2C_{PG})\). With positive quadratic
coefficient, solve the first-order condition and project onto \([0,1]\).

\begin{equation}
\label{eq:26}
w^*=
\min\!\left\{1,\max\!\left\{0,
\frac{V_P-C_{PG}}{V_P+V_G+b^2-2C_{PG}}
\right\}\right\}.
\end{equation}

For \(V_P=0.0025\), \(V_G=0.0001\), \(b=0.10\), and \(C_{PG}=0\), the
optimal weight is \(0.0025/0.0126\approx0.20\). Using inverse variance
alone would give \(0.0025/0.0026\approx0.96\). This difference follows
from the squared bias. Applied weighting requires an estimate or bound
for the bias, and covariance estimation when the estimators share
respondents, training data, or auxiliary information. Account for extra
uncertainty from estimating these quantities.

\subsection{Coverage across tasks and for a fixed
item}\label{coverage-across-tasks-and-for-a-fixed-item}

Let \(T\) index a randomly drawn task and \(U\) the repeated sampling,
correction, or generation process for that task. Let \(I(T,U)\) be the
resulting interval and \(C(t)=\Pr_U\{\theta(t)\in I(t,U)\mid T=t\}\).
The law of total probability gives

\begin{equation}
\label{eq:27}
\overline C=\Pr\{\theta(T)\in I(T,U)\}=E_T\{C(T)\}.
\end{equation}

If task probabilities are \(0.90\) and \(0.10\) and their respective
conditional coverages are \(0.99\) and \(0.59\), the average is \(0.95\)
even though one group has coverage \(0.59\). Huang et al.
\citeyearpar{huang} calibrate for average coverage over a task
distribution. For a fixed questionnaire item and target population, the
assessment must repeat human sampling and the full interval-construction
procedure. The related distinction between marginal and conditional
coverage in predictive inference illustrates why finer guarantees
require additional evidence \citep{barbercond}.

With a known fixed \(\theta\) and \(B\) independent repetitions,
estimate coverage by the fraction of intervals containing \(\theta\).
Its Monte Carlo standard error can be estimated by

\begin{equation}
\label{eq:28}
\widehat C=\frac1B\sum_{b=1}^B\mathbf 1\{\theta\in I_b\},
\qquad \widehat{\operatorname{se}}_{\mathrm{MC}}(\widehat C)
=\sqrt{\frac{\widehat C(1-\widehat C)}{B}}.
\end{equation}

For estimated coverage \(0.95\) and \(B=2000\), the Monte Carlo standard
error is approximately \(0.0049\), or \(0.49\) percentage points. A
coverage study should specify whether the benchmark is a known truth, a
finite-population target, or an estimate itself subject to sampling
error.

\section{Evidence synthesis and benchmark
reanalysis}\label{evidence-synthesis-and-benchmark-reanalysis}

\subsection{Study identification and evidence
coding}\label{study-identification-and-evidence-coding}

We organize the evidence around three questions: which individual,
marginal, and subgroup patterns generated data reproduce, how generation
error affects associations, parameter estimates, and coverage, and which
corrections improve identification or precision when human outcomes are
available. Empirical reports had to compare LLM-generated responses,
distributions, treatment effects, or auxiliary estimators with human
questionnaires, measurements, or behavioral experiments under the same
or a comparable task. Work focused on text generation, model knowledge,
or annotation belongs to neighboring topics. Reviews and statistical
methods publications provide concepts and arguments.

Search and screening ran through November 22, 2026. The candidate set
combined an earlier bibliography of 88 records, the top 20
relevance-ranked results from each of three Crossref queries, and 15
records from hand searching and citation checks. The search window began
in 2022. We used titles, DOIs, and version relationships to merge
duplicates, preprints, and publications. Table~\ref{tab:3} reports the screening
totals. For full-text-verified reports with a human benchmark, we
recorded the benchmark, model inputs, holdout design, baseline,
uncertainty, and main findings. The publication status distinguishes
journal papers, conference papers, preprints, and institutional reports.
PRISMA 2020 informed the reporting of inclusion and exclusion
\citep{page}. The replication materials provide queries,
ranked results, screening decisions, and full-text sources.

\begingroup\setstretch{1}
\begin{longtable}[]{@{}
  >{\raggedright\arraybackslash}p{(\columnwidth - 4\tabcolsep) * \real{0.3333}}
  >{\raggedleft\arraybackslash}p{(\columnwidth - 4\tabcolsep) * \real{0.3333}}
  >{\raggedright\arraybackslash}p{(\columnwidth - 4\tabcolsep) * \real{0.3333}}@{}}
\caption{Candidate records and screening decisions}\label{tab:3}\\
\toprule\noalign{}
\begin{minipage}[b]{\linewidth}\raggedright
Stage or disposition
\end{minipage} & \begin{minipage}[b]{\linewidth}\raggedleft
Records
\end{minipage} & \begin{minipage}[b]{\linewidth}\raggedright
Rule
\end{minipage} \\
\midrule\noalign{}
\endfirsthead
\toprule\noalign{}
\begin{minipage}[b]{\linewidth}\raggedright
Stage or disposition
\end{minipage} & \begin{minipage}[b]{\linewidth}\raggedleft
Records
\end{minipage} & \begin{minipage}[b]{\linewidth}\raggedright
Rule
\end{minipage} \\
\midrule\noalign{}
\endhead
\bottomrule\noalign{}
\endlastfoot
Earlier bibliography & 88 & Reassess relevance to the review question \\
Crossref search & 60 & Retain the first 20 results from each of three
queries \\
Hand search and citation checks & 15 & Add confirmed reports and
candidates awaiting verification \\
Records after deduplication & 156 & Merge versions and remove repeat
hits \\
Included in the empirical synthesis & 27 & Comparable human reference
and completed full-text extraction \\
Background or methods & 40 & Context, theory, or related methods \\
Outside scope & 66 & Task, output, or reference does not meet
eligibility \\
Awaiting verification & 23 & Full text, human reference, or comparison
design still requires verification \\
\end{longtable}
\endgroup

Table~\ref{tab:4} counts reports, and a report can contribute to more than one
level. Studies that reuse the same benchmark share some evidence. Four
design features guide interpretation. The \textbf{human benchmark}
defines the population, survey subset, or experimental participants
under study. \textbf{Information separation} distinguishes
within-archive replication, held-out items, external waves, and
prospective validation. The \textbf{baseline} indicates whether
conventional methods have access to the same inputs. \textbf{Uncertainty
reporting} distinguishes point error, finite-sample confidence
intervals, and coverage under a specified repetition. Matched archive
comparisons describe a particular benchmark. Held-out items or
independent waves assess specified forms of transport. Repetition
compatible with the survey design tests fixed-estimand coverage.

\subsection{Evidence at six levels}\label{evidence-at-six-levels}

\begingroup\setstretch{1}
\begin{longtable}[]{@{}
  >{\raggedright\arraybackslash}p{(\columnwidth - 6\tabcolsep) * \real{0.2500}}
  >{\raggedleft\arraybackslash}p{(\columnwidth - 6\tabcolsep) * \real{0.2500}}
  >{\raggedright\arraybackslash}p{(\columnwidth - 6\tabcolsep) * \real{0.2500}}
  >{\raggedright\arraybackslash}p{(\columnwidth - 6\tabcolsep) * \real{0.2500}}@{}}
\caption{Evaluation levels, benchmarks, and supported conclusions}\label{tab:4}\\
\toprule\noalign{}
\begin{minipage}[b]{\linewidth}\raggedright
Evaluation level
\end{minipage} & \begin{minipage}[b]{\linewidth}\raggedleft
Reports
\end{minipage} & \begin{minipage}[b]{\linewidth}\raggedright
Representative benchmarks and measures
\end{minipage} & \begin{minipage}[b]{\linewidth}\raggedright
Supported conclusion
\end{minipage} \\
\midrule\noalign{}
\endfirsthead
\toprule\noalign{}
\begin{minipage}[b]{\linewidth}\raggedright
Evaluation level
\end{minipage} & \begin{minipage}[b]{\linewidth}\raggedleft
Reports
\end{minipage} & \begin{minipage}[b]{\linewidth}\raggedright
Representative benchmarks and measures
\end{minipage} & \begin{minipage}[b]{\linewidth}\raggedright
Supported conclusion
\end{minipage} \\
\midrule\noalign{}
\endhead
\bottomrule\noalign{}
\endlastfoot
Individual responses & 6 & Matched profiles, held-out items, agreement,
and human retests & Richer profiles help on some items. The information
available at deployment governs transport. \\
Marginal distributions & 23 & ANES, Pew, WVS, GSS, and CGSS. Option
shares, means, and distribution distances & Some population
distributions can be approximated. Error varies by item, wave, and
protocol, requiring external evaluation in the intended population. \\
Joint structure & 5 & ANES, CGSS, Eurobarometer, and SSDataBench.
Covariances, regressions, and life-event sequences & Similar margins can
coexist with distorted associations. External checks of transitions and
dependence remain limited. \\
Subgroup contrasts & 19 & US surveys, Global Preference Survey, and
cross-cultural instruments. Differences, signs, and ranks & Fidelity
varies by demographic, country, and cultural dimension. Small groups,
cross-cultural transport, and small contrasts require separate
evaluation. \\
Parameter estimation & 7 & Human treatment effects, direct survey means,
and repeated sample benchmarks & Correction with human outcomes can
improve some means. Direction, rank, and magnitude of treatment effects
require distinct measures. \\
Inferential validity & 2 & OpinionQA, NHANES, and the American Trends
Panel. Coverage, width, and effective sample size & Evidence concerns
coverage across tasks and selected empirical mean-estimation settings.
Fixed items and small domains lack cross-benchmark coverage evidence. \\
\end{longtable}
\endgroup

\emph{Note.} Counts overlap. Parameter estimation includes correction of
survey estimands and treatment-effect estimation relative to human
experiments. Regression and covariance fidelity enter joint structure.
Confidence intervals and difference tests can enter other levels;
studies that test interval coverage enter inferential validity.

\textbf{Individual prediction depends on the information available at
deployment.} Park et al. \citeyearpar{park} report that adding survey
and interview information increases held-out GSS item accuracy from
60.4\% to 72.8\%, compared with 79.5\% human retest agreement. The gain
concerns predictions for a person whose other information was available.
D'Ecclesiis et al. \citeyearpar{decclesiis} find that some models
reproduce aggregate and subgroup patterns despite limited individual
agreement in Eurobarometer data. Ahnert et al. \citeyearpar{ahnert}
compare elicitation with the same model and find that constrained
generation generally performs well on the closed-ended political items
studied. Imputation assessment requires a baseline with the same inputs
and holdouts at the relevant respondent and outcome levels. Accuracy
comparisons without human retests combine model performance and
differences in item repeatability.

\textbf{Marginal error varies across benchmarks and protocols.} Li et
al. \citeyearpar{li} find that aggregate share predictions outperform
uniform, normal, and historical-average baselines on several Pew surveys
while retaining partisan discrepancies. Chapekis et al.
\citeyearpar{pew} reproduce three recent American Trends Panel
questionnaires and report an average absolute share error of roughly 12
percentage points across nearly 300 questions. The studies use different
waves, topics, inputs, and metrics. The first demonstrates gains over
particular baselines, and the latter shows that absolute error remains
consequential on newer topics even with richer profiles. Survey-specific
training improved some held-out predictions in Suh et al.
\citeyearpar{suh}. Historical calibration improved multiple held-out
items in Jiang et al. \citeyearpar{jiang}, with variable gains over a
human-history-only baseline. Evaluations should report absolute error,
information-matched baselines, and a prespecified tolerance.

\textbf{Test dispersion, association, and longitudinal transitions
separately.} In the ANES feeling thermometer analysis of Bisbee et al.
\citeyearpar{bisbee}, 48\% of generated-data regression coefficients
differed significantly from their human counterparts despite relatively
close aggregate means. Cui and Xiong \citeyearpar{cuixiong} analyze ten
common attitude items over two waves of the Chinese General Social
Survey. Some income, urban--rural, and sex coefficients were distorted,
and changing the standardization changed some test decisions. Across
seven cross-sectional or longitudinal benchmarks, Xie et al.
\citeyearpar{xie} report near-zero pass rates for tests of life-event
sequence distributions. These examples illustrate the mechanisms in
Sections 5.1 and 5.2: cancellation, moment discrepancies, and temporal
dependence. Paired statistics require a common coding and measurement
scale. Pass rates also depend on sample size and test power, so they
should be interpreted alongside error magnitudes and intervals.

\textbf{Transport of subgroup contrasts depends on conditional response
fidelity.} Santurkar et al. \citeyearpar{santurkar} find persistent
mismatches after conditioning on US subgroup identity. Across 74
countries in the Global Preference Survey, Chen et al.
\citeyearpar{chen} report, under their interval-based criterion, 77.8\%
agreement on demographic associations and 48.1\% on cultural dimensions.
Durmus et al. \citeyearpar{durmus} show that specifying a country
perspective and translating the item yield different alignment patterns.
AlKhamissi et al. \citeyearpar{alkhamissi} find larger discrepancies for
underrepresented profiles or culturally sensitive items in Egypt and the
United States. Qu and Wang \citeyearpar{qu} also document variation by
country and demographics. In open-ended tasks, Wang et al.
\citeyearpar{wangidentity} compare model replies with those of group
members and find that identity prompting can reduce within-group
diversity. These results motivate direct evaluation for the intended
group and item. Cross-country averages provide limited evidence for
underrepresented countries or languages.

\textbf{Experiment effects and survey means require different human
checks.} In the main US benchmark of Ashokkumar et al.
\citeyearpar{hewitt}, GPT-4 forecasts correlate 0.85 with observed
effects, with a calibration slope of 0.56. The pooled correlation in the
secondary megastudies is 0.39. Cui et al. \citeyearpar{cui} report that
three models reproduce 73\%--81\% of main effects and 46\%--63\% of
interactions in scenario-based psychology and management experiments. Wu
et al. \citeyearpar{wu} replicate some information-treatment effects on
inflation expectations, with results depending on historical cutoffs and
prompt information. The classic behavioral experiments of Aher et al.
\citeyearpar{aher} and the request game of Gao et al. \citeyearpar{gao}
show that some behavioral patterns are recoverable while some decision
distributions differ substantially. These results support candidate
screening under validated conditions. Sample planning and
decision-making also require relevant human experiments to establish the
scale and prediction error. For one dietary intake mean, tuned residual
correction increased effective sample size by about 14\%, whereas
unit-weight correction lowered precision \citep{krsteski}. The residual
variance conditions in Section 5.4 explain this dependence on tuning.

\textbf{Inferential validity depends on the unit of repetition.} Huang
et al. \citeyearpar{huang} use human calibration items and target
coverage for new items averaged over the task distribution. Krsteski et
al. \citeyearpar{krsteski} assess empirical coverage for corrected
estimates under a specified human benchmark and task. These designs
address across-item average coverage and fixed-task coverage under
repeated sampling, respectively. The former requires transfer checks by
construct, language, and subgroup. The latter requires repetition that
follows the human design and incorporates appropriate weights.
Generation variance and interval width alone cannot establish either
form of coverage.

\subsection{Paired effect reanalysis of two public
benchmarks}\label{paired-effect-reanalysis-of-two-public-benchmarks}

\subsubsection{Analysis units, inclusion, and
estimators}\label{analysis-units-inclusion-and-estimators}

Two public benchmark datasets supply paired GPT-4 and human point
estimates with identifiers for the original papers or experiments
\citep{cui, hewitt}. The main analysis concerns these archives and model
configurations. The Cui archive records effects on a correlation scale.
Its primary analysis comprises 578 directional main-effect pairs from
148 experiments in 49 papers. Another 83 pairs have both effect
estimates, of which 81 are omnibus effects and two lack clear direction.
We include these in an expanded sensitivity analysis. The Ashokkumar
archive uses all pairwise treatment-condition comparisons from Figure
1a. Removing mirrored comparisons with the treatment and reference
conditions swapped leaves 1,678 pairs from 70 experiments. The primary
result in the original study uses a reference condition and outcome
sampled per experiment. Our sensitivity analysis gives outcomes equal
weight within each experiment.

Let \(g_{aki}\) and \(h_{aki}\) be the generated and human point
estimates in a given pair. The pair's amplification score is

\begin{equation}
\label{eq:29}
s_{aki}=\mathbf 1\{|g_{aki}|>|h_{aki}|\}
+\tfrac12\mathbf 1\{|g_{aki}|=|h_{aki}|\}.
\end{equation}

Let \(K_a\) denote the number of clusters in archive \(a\), and let
\(n_{ak}\) be the number of pairs in cluster \(k\). We average within
clusters, give clusters equal weight within archives, and give each of
the two archives one-half of the combined weight.

\begin{equation}
\label{eq:30}
\widehat p=\frac{1}{2}\sum_{a=1}^{2}\frac{1}{K_a}\sum_{k=1}^{K_a}\frac{1}{n_{ak}}\sum_{i=1}^{n_{ak}}s_{aki}.
\end{equation}

We use 20,000 stratified cluster bootstrap samples at the original study
level and report percentile intervals \citep{field}. The combined
frequency is scale-free, while amplification magnitude is reported on
each archive's own scale. For magnitude, we take each cluster's median
of the difference in absolute effects and summarize those medians across
clusters. The bootstrap treats published pair-level estimates as inputs.
For the Cui archive, we also transform the reported human effect
intervals into absolute-value ranges and tally fixed model estimates
above, below, and within those ranges.

\subsubsection{Main results and sensitivity
checks}\label{main-results-and-sensitivity-checks}

The main analysis contains 2,256 paired effects and 119 clusters. The
two unrounded archive rates in~\eqref{eq:30} are approximately 0.68447 and
0.52868, giving \((0.68447+0.52868)/2\approx0.60658\). Giving the
clusters equal weight across both archives gives
\((49\times0.68447+70\times0.52868)/119\approx0.59283\). Under equal
archive weights, the amplification rate is \textbf{60.7\%} (95\% cluster
bootstrap interval \textbf{54.9\%--66.2\%}). The scenario archive gives
68.4\% and the US text-survey archive gives 52.9\%. Their difference is
15.6 percentage points (4.1--26.9). Study tasks, effect scales, and
measurement protocols vary jointly between the two archives, so the
difference describes these archives. Figure~\ref{fig:5} gives the overall and
subgroup results.

\begingroup\setstretch{1}
\begin{figure}
\centering
\includegraphics[width=0.98\textwidth,height=\textheight]{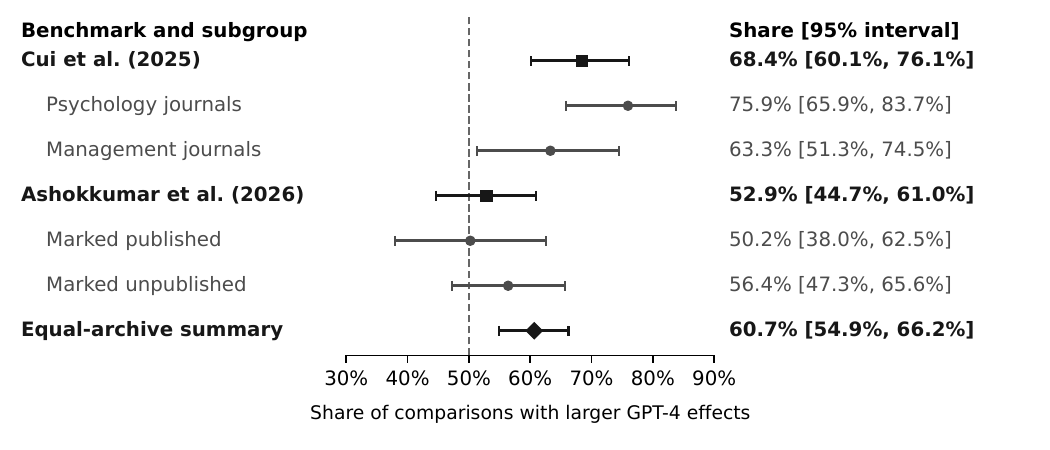}
\caption{Frequency of GPT-4 effect amplification. The dashed line marks
equal frequency, the diamond shows the equal-archive estimate, and
horizontal bars show 95\% within-archive cluster bootstrap intervals.}
\label{fig:5}
\end{figure}
\endgroup

\begingroup\setstretch{1}
\begin{longtable}[]{@{}
  >{\raggedright\arraybackslash}p{(\columnwidth - 4\tabcolsep) * \real{0.3333}}
  >{\raggedright\arraybackslash}p{(\columnwidth - 4\tabcolsep) * \real{0.3333}}
  >{\raggedright\arraybackslash}p{(\columnwidth - 4\tabcolsep) * \real{0.3333}}@{}}
\caption{Effect amplification frequency and magnitude by archive}\label{tab:5}\\
\toprule\noalign{}
\begin{minipage}[b]{\linewidth}\raggedright
Archive
\end{minipage} & \begin{minipage}[b]{\linewidth}\raggedright
Frequency and 95\% cluster bootstrap interval
\end{minipage} & \begin{minipage}[b]{\linewidth}\raggedright
Median absolute-effect difference {[}interquartile range{]}
\end{minipage} \\
\midrule\noalign{}
\endfirsthead
\toprule\noalign{}
\begin{minipage}[b]{\linewidth}\raggedright
Archive
\end{minipage} & \begin{minipage}[b]{\linewidth}\raggedright
Frequency and 95\% cluster bootstrap interval
\end{minipage} & \begin{minipage}[b]{\linewidth}\raggedright
Median absolute-effect difference {[}interquartile range{]}
\end{minipage} \\
\midrule\noalign{}
\endhead
\bottomrule\noalign{}
\endlastfoot
Psychology and management scenario experiments & 68.4\% {[}60.1\%,
76.1\%{]} & 0.150 {[}0.044, 0.283{]}, correlation scale \\
US text-based survey experiments & 52.9\% {[}44.7\%, 61.0\%{]} & 0.66
{[}\ensuremath{-}1.04, 2.51{]} percentage points \\
Equal-archive summary & 60.7\% {[}54.9\%, 66.2\%{]} & Archive-specific
scales \\
\end{longtable}
\endgroup

Expanding the Cui archive to all pairs with both estimates raises the
combined rate to 61.1\%. Equal outcome weights within Ashokkumar
experiments leave the point estimate at 52.9\%. Requiring at least 10\%
amplification produces a combined rate of 58.1\%. At a 25\% threshold,
the combined interval includes 50\%. In the Cui archive, 51.8\% are
above the upper bound of the absolute-value interval for the human
effect, 13.8\% below the lower bound, and 34.4\% within the interval.
This diagnostic shows how often the comparison holds for all human
effects in the published interval, with the generated point estimate
fixed. The Ashokkumar Figure 1a data provide no pair-level intervals or
covariances, so we report the diagnostic separately from the two-archive
point-estimate comparison.

\begingroup\setstretch{1}
\begin{longtable}[]{@{}
  >{\raggedright\arraybackslash}p{(\columnwidth - 6\tabcolsep) * \real{0.2500}}
  >{\raggedleft\arraybackslash}p{(\columnwidth - 6\tabcolsep) * \real{0.2500}}
  >{\raggedleft\arraybackslash}p{(\columnwidth - 6\tabcolsep) * \real{0.2500}}
  >{\raggedright\arraybackslash}p{(\columnwidth - 6\tabcolsep) * \real{0.2500}}@{}}
\caption{Sensitivity to inclusion, weights, and amplification thresholds}\label{tab:6}\\
\toprule\noalign{}
\begin{minipage}[b]{\linewidth}\raggedright
Analysis
\end{minipage} & \begin{minipage}[b]{\linewidth}\raggedleft
Cui archive
\end{minipage} & \begin{minipage}[b]{\linewidth}\raggedleft
Ashokkumar archive
\end{minipage} & \begin{minipage}[b]{\linewidth}\raggedright
Summary and 95\% cluster bootstrap interval
\end{minipage} \\
\midrule\noalign{}
\endfirsthead
\toprule\noalign{}
\begin{minipage}[b]{\linewidth}\raggedright
Analysis
\end{minipage} & \begin{minipage}[b]{\linewidth}\raggedleft
Cui archive
\end{minipage} & \begin{minipage}[b]{\linewidth}\raggedleft
Ashokkumar archive
\end{minipage} & \begin{minipage}[b]{\linewidth}\raggedright
Summary and 95\% cluster bootstrap interval
\end{minipage} \\
\midrule\noalign{}
\endhead
\bottomrule\noalign{}
\endlastfoot
Main analysis, equal archive weights & 68.4\% & 52.9\% & 60.7\%
{[}54.9\%, 66.2\%{]} \\
Equal weights for 119 original-study clusters & 68.4\% & 52.9\% & 59.3\%
{[}53.4\%, 65.0\%{]} \\
Equal weights for outcomes within Ashokkumar experiments & 68.4\% &
52.9\% & 60.7\% {[}54.8\%, 66.2\%{]} \\
Include 83 additional pairs in the Cui archive & 69.4\% & 52.9\% &
61.1\% {[}55.3\%, 66.7\%{]} \\
Restrict to matching human and generated directions & 71.1\% & 57.5\% &
64.3\% {[}58.0\%, 70.4\%{]} \\
More than 10\% above the absolute human estimate & 65.6\% & 50.6\% &
58.1\% {[}52.3\%, 63.7\%{]} \\
More than 25\% above the absolute human estimate & 62.8\% & 48.2\% &
55.5\% {[}49.6\%, 61.2\%{]} \\
\end{longtable}
\endgroup

Amplification occurs more often in the scenario replications. The
text-based experiment interval includes the equal-frequency reference.
Both archives show amplification even among pairs with the same
direction. Decisions using effect magnitudes require scale calibration,
accounting for benchmark uncertainty, and external validation. Extension
to other models, non-text interventions, or populations requires
additional paired data.

\section{Principles for practice and future research}\label{practical-implications-and-research-agenda}

The evidence supports three principles for using synthetic survey data. Specify the target population and estimand before choosing a generation protocol. Document what each human data source contributes to estimation, including the sampling or selection mechanism that links it to the target. Match the reported uncertainty to the source of variation relevant to the decision. The choice of reference data then follows from the intended use. Questionnaire piloting calls for cognitive validation, subgroup comparisons require group-specific benchmarks, and population intervals require a design or model that supports inference about human outcomes.

\textbf{Set the estimand and the human-data budget first.} Generated
replies can flag ambiguous wording in an instrument pilot, with
cognitive interviews used to investigate the problem in human
respondents. Aggregate and subgroup forecasts should report external
error in the deployment population. Population estimates and confidence
intervals require a stated sampling design and a basis for estimating
residuals. For subgroup contrasts, report precision within both groups
and the error and uncertainty of the contrast. When the aim is to
establish that the generated statistic is sufficiently close to the
human statistic, prespecify a tolerance \(\Delta>0\). With \(D\) the
population difference, equivalence tests compare

\begin{equation}
\label{eq:31}
H_0:\ D\le-\Delta\ \text{or}\ D\ge\Delta,
\qquad H_1:\ -\Delta<D<\Delta.
\end{equation}

For two one-sided tests of size \(\alpha\), a compatible \((1-2\alpha)\)
confidence interval must lie entirely in \((-\Delta,\Delta)\) to support
equivalence \citep{lakens}. When reusing respondents, items, or
experiments in paired comparisons, account for the dependence in the
standard error.

\textbf{Vary the available inputs and target settings separately.}
Benchmark designs should hold out respondents and new items, incorporate
later waves, and match translated instruments with appropriate human
benchmarks. Holding out respondents under a fixed instrument assesses
prediction for new respondents. Holding out item families addresses
transfer to different constructs. Holding out later waves tests temporal
transfer. Factorial variation in model version, profile detail, and
elicitation can estimate their effects and interactions. Define separate
data partitions for training, prompt development, and final evaluation.

\textbf{Compare direct and model-assisted estimation at equal cost.}
Define the outcome, target population, sampling design, and human-data
costs. Randomly vary the fraction of labeled outcomes assigned to model
adaptation and to correction. With matched information, compare direct
survey estimates, design-weighted regression adjustment with
conventional predictors, and LLM-assisted estimates. Report design bias,
mean squared error, coverage, and total cost. For small groups, specify
sampling allocations and check the weight distribution. Record failures
and refusals before restricting analysis to complete outputs.

\textbf{Limit profile information to the scope of consent and the
research purpose.} Detailed profiles can contain identifiers or permit
identification through attribute combinations. Before sending data to an
API, restrict access, minimize data use, and check for leakage. To make
the study reproducible, retain the instrument version, model identifier,
date, prompt template, decoding settings, parsing procedure, and
analysis rules. Where consent limits publication of records or outputs,
release reproducible summaries or demonstrate the processing with
synthetic test cases. When changing models, revalidate analysis outcomes
under a prespecified design.

\begingroup\setstretch{1}
\begin{longtable}[]{@{}
  >{\raggedright\arraybackslash}p{(\columnwidth - 4\tabcolsep) * \real{0.3333}}
  >{\raggedright\arraybackslash}p{(\columnwidth - 4\tabcolsep) * \real{0.3333}}
  >{\raggedright\arraybackslash}p{(\columnwidth - 4\tabcolsep) * \real{0.3333}}@{}}
\caption{Research questions and testable designs}\label{tab:7}\\
\toprule\noalign{}
\begin{minipage}[b]{\linewidth}\raggedright
Question
\end{minipage} & \begin{minipage}[b]{\linewidth}\raggedright
Current evidence
\end{minipage} & \begin{minipage}[b]{\linewidth}\raggedright
Proposed design and criterion
\end{minipage} \\
\midrule\noalign{}
\endfirsthead
\toprule\noalign{}
\begin{minipage}[b]{\linewidth}\raggedright
Question
\end{minipage} & \begin{minipage}[b]{\linewidth}\raggedright
Current evidence
\end{minipage} & \begin{minipage}[b]{\linewidth}\raggedright
Proposed design and criterion
\end{minipage} \\
\midrule\noalign{}
\endhead
\bottomrule\noalign{}
\endlastfoot
Target population change & Cross-country preferences and partisan
differences can be distorted & Hold out countries and later waves.
Compare weighted subgroup differences against fixed tolerances. \\
Distorted dependence between items & Similar margins can coexist with
different joint distributions or regressions & Compare joint with
itemwise elicitation for matched profiles. Check residual covariance and
scale reliability. \\
Allocation of human outcomes & Prediction quality and gains from
residual correction vary by task & Randomize label allocations at equal
cost. Compare design coverage and mean squared error. \\
Transport of uncertainty statements & Across-task and fixed-estimand
coverage use different forms of repetition & Assess coverage separately
over held-out item families and repeated human samples. \\
Information leakage and model version & Profile matching and replication
may use results seen during model development & Record text provenance.
Use prospective items and compare model versions under fixed prompts. \\
\end{longtable}
\endgroup

Table~\ref{tab:7} and Figure~\ref{fig:6} connect research questions with validation designs
and outcomes.

\begingroup\setstretch{1}
\begin{figure}
\centering
\includegraphics[width=0.94\textwidth,height=\textheight]{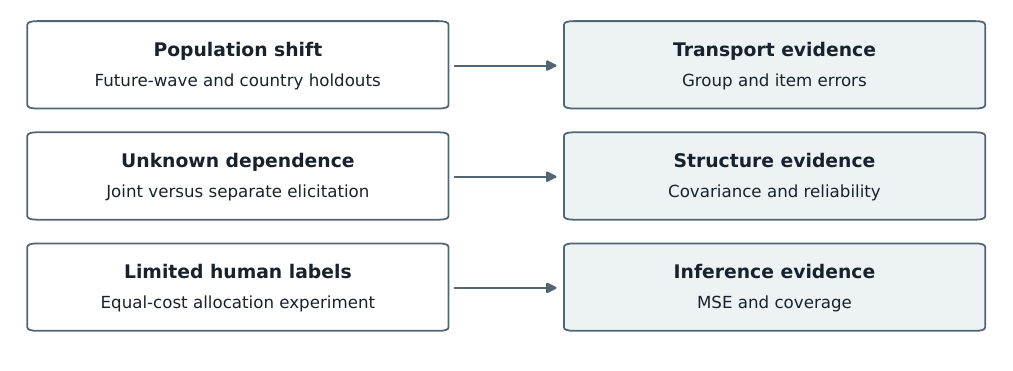}
\caption{Research designs and their validation targets.}
\label{fig:6}
\end{figure}
\endgroup

\section{Conclusion}\label{conclusion}

The statistical use of LLM-generated survey samples depends on the
estimand, available input information, and human reference data. The
evidence is most extensive for marginal distributions and subgroup
contrasts. Some public-opinion proportions can be approximated, while
joint associations, cultural contrasts, and longitudinal trajectories
show weaker stability. Agreement at the overall mean constrains one
statistic. Cancellation of subgroup errors and divergence in joint
structure can change the results of secondary analyses.

Generation and population inference involve different sources of
uncertainty. Further draws narrow Monte Carlo error under the generation
law. Residual correction with a human probability sample identifies a
population mean under a specified sampling design, and the gain in
precision depends on residual variation. Few studies investigate
interval coverage, and their units of repetition differ. For a fixed
item or a small subgroup, assess coverage under repetition that matches
the intended design.

The paired reanalysis finds differences in the frequency and magnitude
of GPT-4 effect amplification across two public benchmarks.
Amplification is more frequent in the scenario archive, and the
text-survey interval includes the equal-frequency reference. Externally
validated predictions can support candidate screening. Sample planning
and policy decisions that depend on effect magnitudes require
calibration against relevant human experiments. Priority validation
settings include new waves, new items, and small groups, with direct and
model-assisted estimators compared under equal human-data budgets.

\section*{Data Availability Statement}

The paired reanalysis uses publicly available effect estimates from the replication archive of \citet{cui} and the Figure 1 source data of \citet{hewitt}. The companion replication archive contains the paired estimates, analysis code, and literature screening records.

\section*{Acknowledgments}

OpenAI's ChatGPT assisted with drafting and editing the manuscript text and formatting the submission files.

\bibliography{references}

\end{document}